# Electrodynamics of Conductive Oxides: Intensity-dependent anisotropy, reconstruction of the effective dielectric constant, and harmonic generation

Michael Scalora,[1] Jose Trull,[2] Domenico de Ceglia,[3] Maria Antonietta Vincenti,[4] Neset Akozbek,[5] Zachary Coppens,[5] Laura Rodríguez-Suné,[2] Crina Cojocaru[2]

[1] *Charles M. Bowden Research Center, CCDC AVMC, Redstone Arsenal, AL 35898-5000 USA*
[2] *Department of Physics, Universitat Politècnica de Catalunya, 08222 Terrassa, Spain*
[3] *Department of Information Engineering – University of Padova, Via Gradenigo 6/a 35131 Padova, Italy*
[4] *Department of Information Engineering – University of Brescia, Via Branze 38 25123 Brescia, Italy*
[5] *AEgis Technologies Inc., 401 Jan Davis Dr., Huntsville, Alabama 35806, USA*

**Abstract:** We study electromagnetic pulse propagation in an indium tin oxide nanolayer in the linear and nonlinear regimes. We use the constitutive relations to reconstruct the effective dielectric constant of the medium, and show that nonlocal effects induce additional absorption resonances and anisotropic dielectric response: longitudinal and transverse effective dielectric functions are modulated differently along the propagation direction, and display different epsilon-near-zero crossing points with a discrepancy that increases with increasing intensity. We predict that hot carriers induce a dynamic redshift of the plasma frequency and a corresponding translation of the effective nonlinear dispersion curves that can be used to predict and quantify nonlinear refractive index changes as a function of incident laser peak power density. Our results suggest that large, nonlinear refractive index changes can occur without the need for epsilon-near-zero modes to couple with plasmonic resonators. At sufficiently large laser pulse intensities, we predict the onset of optical bistability, while the presence of additional pump absorption resonances that arise from longitudinal oscillations of the free electron gas give way to corresponding resonances in the second and third harmonic spectra. A realistic propagation model is key to unraveling the basic physical mechanisms that play a fundamental role in the dynamics.

## 1. Introduction

Typical plasmonic resonators consist of metallic nanoparticles or nanostructures where free electrons oscillate in resonance with light. These resonances can produce strong field amplification and enhanced scattering/absorption cross sections, which are key properties for applications in sensing, detection, energy harvesting and generic light manipulation at the nanoscale. However, metals can be either too absorptive or inadequate in a given wavelength range, and alternative replacements must be sought. In this work we explore novel linear and nonlinear propagation effects that manifest themselves but are not limited to free electron systems that may display an epsilon-near-zero (or ENZ) crossing of the real part of the dielectric constant. In particular, we study the basic properties of simple layers composed of degenerate semiconductors like Indium Tin Oxide (ITO) only a few tens of nanometers in thickness in order to ascertain basic physical characteristics that may transfer to more complicated nano-structured geometries. Generally, free electron systems are centrosymmetric and are described by a simplistic Drude model. However, experiments show that in reality these materials possess a combined Lorentz-Drude-like dielectric response [1] that can be tuned by controlling doping levels and annealing temperatures. This dual material aspect simultaneously complicates and enriches the dynamics, whose understanding and description thus require theoretical models that are more comprehensive than what may be required in ordinary photonic structures.

In contrast to noble metals, conducting oxides display lower losses and may thus substitute or even supplant metals in certain applications and spectral wavelength ranges. To date, many aspects related to pulse propagation phenomena in free electron systems like noble metals or conducting oxides remain incomplete. In what follows we describe a model that simultaneously accounts for: *(i)* the intrinsic nonlinearities of background bound charges; *(ii)* nonlocal effects (pressure and viscosity of the electron gas); *(iii)* pump depletion; *(iv)* the dynamics that ensue from including an intrinsic, temperature-dependent effective mass (in the case of conducting oxides) or free charge density (in the case of noble metals or semiconductors) and related nonlinearities that ultimately manifest themselves in the form of effective $\chi^{(3)}$, $\chi^{(5)}$, and higher order nonlinear contributions; and *(v)* surface and magnetic nonlinearities that are almost always neglected in favor of bulk nonlinearities. As an example of this theoretical deficiency in conducting oxides, and to some extent in metals and semiconductors, the nature and magnitude of nonlinear index of refraction changes as a function of incident pump intensity has not yet been clarified [2,3]. Differing explanations have been provided regarding the source of third order phenomena [3, 4], and practically no good insight into second order, surface and magnetic phenomena outside of the context found in references [1] and [5]. In reference [3] third order phenomena responsible for nonlinear index changes were attributed to the free electron cloud. In reference [4] third harmonic generation was attributed exclusively to the background crystal. In reference [6], THG from an ITO nanolayer was studied theoretically and experimentally using an generic, dispersionless $\chi^{(3)}$ having no specified origin. Finally, in reference [7], the simultaneous generation of negatively refracted and phase conjugate beams was experimentally recorded from a structure consisting of gold nanoantennas patterned on top of a 40nm-thick ITO layer displaying an ENZ crossing point. However, the theoretical effort tackled only generic aspects of a third order nonlinearity present only in the ITO layer, and no detailed field dynamics. In summary, the picture that emerges from the detailed microscopic model discussed below is somewhat more complicated than it would appear in references [3], [4], and [7].

In a recent paper [1] experimental and theoretical results on second and third harmonic generation (SHG and THG) were reported near the ENZ condition of an ITO nanolayer, which manifested itself near 1240nm. The pulse propagation model that was used comprised a hydrodynamic description of the material equations that takes into account free and bound charges, nonlocal effects, a time-dependent free-electron plasma frequency, surface, magnetic, and convective second and third order nonlinearities, as well as the inclusion of linear and nonlinear contributions of the background medium to the dielectric constant. A direct comparison of the SHG spectra and the angular dependence of SH conversion efficiencies showed good qualitative and quantitative agreement with experimental results. Good qualitative and quantitative agreement was also found for the angular dependence of THG for incident laser pulse power densities in the 1 GW/cm$^2$ range. In our present effort we provide further details about the model by: (*i)* expanding the range of investigation well into the IR range; (*ii)* examining the linear regime in order to ascertain the multifaceted contributions of nonlocal effects; and (*iii)* extending our predictions into the high intensity regime in an attempt to distinguish between bound and free (hot) electron contributions.

The local dielectric constant of any material may be expressed as a superposition of Lorentz and Drude oscillators, which in the simplest case of two polarization species (one free and one bound electron contributions, as in the case of ITO [1]) may be written as follows:

$$\varepsilon_{ITO}(\omega) = 1 - \frac{\omega_p^2}{\omega^2 + i\gamma_f \omega} - \frac{\omega_{p,b}^2}{\omega^2 - \omega_{0,b}^2 + i\gamma_b \omega}, \qquad (1)$$

where $\omega_p^2 = \frac{4\pi n_{0f} e^2}{m_f^*}$ is the free electron plasma frequency; $n_{0f}$ the free electron density; $e$ the electronic charge; $\gamma_f$ the free electron damping coefficient; $m_f^*$ the effective free electron mass; $\omega_{p,b}$ is the bound electron plasma frequency, defined similarly to the free electron counterpart to which there corresponds a bound electron density $n_b$ and mass $m_b^*$; $\omega_{0,b}$ is the resonance frequency; and $\gamma_b$ the bound electron damping coefficient.

If the free-electron effective mass changes approximately linearly with temperature, as shown below using the two-temperature model, the equations of motion are reproduced from reference [1], and may be written as follows:

$$\begin{aligned}\ddot{\mathbf{P}}_f + \tilde{\gamma}_f \dot{\mathbf{P}}_f = {} & \frac{n_{0,f} e^2 \lambda_0^2}{m_0^* c^2}\mathbf{E} - \frac{e\,\lambda_0}{m_0^* c^2}\left(\nabla \bullet \mathbf{P}_f\right)\mathbf{E} + \left(\sum_{l=1,2,3}\left(-\tilde{\Lambda}\right)^l \left(\mathbf{E}\bullet\mathbf{E}\right)^l\right)\mathbf{E} + \frac{e\,\lambda_0}{m_0^* c^2}\dot{\mathbf{P}}_f \times \mathbf{H} \\ & + \frac{3E_F}{5m_0^* c^2}\left(\nabla\left(\nabla\bullet\mathbf{P}_f\right) + \frac{1}{2}\nabla^2\mathbf{P}_f\right) - \frac{1}{n_{0,f} e\lambda_0}\left[\left(\nabla\bullet\dot{\mathbf{P}}_f\right)\dot{\mathbf{P}}_f + \left(\dot{\mathbf{P}}_f\bullet\nabla\right)\dot{\mathbf{P}}_f\right]\end{aligned}, \qquad (2)$$

$$\ddot{\mathbf{P}}_b + \tilde{\gamma}_b \dot{\mathbf{P}}_b + \tilde{\omega}_{0,b}^2 \mathbf{P}_b + \mathbf{P}_{b,NL} = \frac{n_{0,b} e^2 \lambda_0^2}{m_b^* c^2}\mathbf{E} + \frac{e\,\lambda_0}{m_b^* c^2}(\mathbf{P}_b \bullet \nabla)\mathbf{E} + \frac{e\,\lambda_0}{m_b^* c^2}\dot{\mathbf{P}}_b \times \mathbf{H}. \qquad (3)$$

Time and space have been scaled such that temporal and spatial derivatives are carried out with respect to the following coordinates: $\varsigma = y/\lambda_0$, $\xi = z/\lambda_0$, and $\tau = ct/\lambda_0$, where in our case $\lambda_0 = 1\mu\text{m}$ is a convenient reference wavelength. It follows that the coefficients are also scaled: $\tilde{\gamma}_{f,b} = \gamma_{f,b}\lambda_0/c$, $\tilde{\omega}_{0,b}^2 = \omega_{0,b}^2\lambda_0^2/c^2$. Eq.(2) describes the free electron polarization, $\mathbf{P}_f$; $\frac{n_{0,f} e^2 \lambda_0^2}{m_0^* c^2}\mathbf{E} - \frac{e\,\lambda_0}{m_0^* c^2}\mathbf{E}\left(\nabla\bullet\mathbf{P}_f\right)$ are Coulomb terms generated by the continuity equation; $\left(\sum_{l=1,2,3}\left(-\tilde{\Lambda}\right)^l \left(\mathbf{E}\bullet\mathbf{E}\right)^l\right)\mathbf{E}$, where $\tilde{\Lambda}$ is a constant of proportionality, follows from the expansion of the effective mass as a function of temperature and absorption [1, 8] (as the summation index indicates, looking ahead to our results our parameter choices demand we retain hot electron nonlinearities up to seventh order;) $\frac{e\,\lambda_0}{m_0^* c^2}\dot{\mathbf{P}}_f \times \mathbf{H}$ arises from the magnetic Lorentz force; $\frac{3E_F}{5m_0^* c^2}\left(\nabla\left(\nabla\bullet\mathbf{P}_f\right) + \frac{1}{2}\nabla^2\mathbf{P}_f\right)$ represent pressure and viscosity, respectively, where $E_F = \frac{\hbar^2}{2m_0^*}\left(3\pi^2 n_{0f}\right)^{2/3}$ is the Fermi energy and we have neglected damping terms [9]; ending with the first order convective contribution $\frac{1}{n_{0,f} e\lambda_0}\left[\left(\nabla\bullet\dot{\mathbf{P}}_f\right)\dot{\mathbf{P}}_f + \left(\dot{\mathbf{P}}_f\bullet\nabla\right)\dot{\mathbf{P}}_f\right]$. Eq.(3) in turn describes the dynamics of bound electrons; $\mathbf{P}_b$ is the bound electrons' polarization; $\mathbf{P}_{b,NL} = \tilde{\alpha}\mathbf{P}_b\mathbf{P}_b - \tilde{\beta}(\mathbf{P}_b\bullet\mathbf{P}_b)\mathbf{P}_b + ....$ is the bound electron's nonlinear polarization, depicted here up to third order; $\frac{n_{0,b}e^2\lambda_0^2}{m_b^* c^2}\mathbf{E}+\frac{e\,\lambda_0}{m_b^* c^2}(\mathbf{P}_b\bullet\nabla)\mathbf{E}$ are Coulomb terms, followed by the magnetic Lorentz term, $\frac{e\lambda_0}{m_b^* c^2}\dot{\mathbf{P}}_b\times\mathbf{H}$. The coefficients $\tilde{\alpha}$ and $\tilde{\beta}$ are tensors that reflect crystal symmetry. In what follows we assume ITO is centrosymmetric ($\tilde{\alpha}=0$) and isotropic, so $\tilde{\beta}$ is a constant. Eqs.(2) and (3) are integrated together with the vector Maxwell equations, where the total polarization is expressed as the vector sum of all polarization components, in this case $\mathbf{P}_{Total}=\mathbf{P}_f+\mathbf{P}_b$.

## 2. Nonlocal Effects

### 2.1 Linear Absorption

At low incident power densities, in a two-dimensional geometry (invariant in the *x*-direction; see Fig.1 for an elucidation of the geometry) only linear, nonlocal effects survive. The free electron component Eq.(2) may then be rewritten as follows:

$$\ddot{\mathbf{P}}_f+\tilde{\gamma}_f\dot{\mathbf{P}}_f=\frac{n_{0,f}e^2\lambda_0^2}{m_0^* c^2}\mathbf{E}+\frac{3E_F}{5m_0^* c^2}\left[\left(\frac{\partial}{\partial y}\hat{\mathbf{j}}+\frac{\partial}{\partial z}\hat{\mathbf{k}}\right)\left(\frac{\partial P_y}{\partial y}+\frac{\partial P_z}{\partial z}\right)+\frac{1}{2}\left(\frac{\partial^2}{\partial y^2}+\frac{\partial^2}{\partial z^2}\right)\left(P_y\hat{\mathbf{j}}+P_z\hat{\mathbf{k}}\right)\right], \quad (4)$$

where $\hat{\mathbf{j}}$ and $\hat{\mathbf{k}}$ are unit vectors along $y$ and $z$, respectively. We continue to assume appropriately scaled Cartesian coordinates, but for clarity we have retained the usual notation. With the spatial derivatives such that $\frac{\partial}{\partial y}\to i\tilde{k}_y$ and $\frac{\partial}{\partial z}\to i\tilde{k}_z$, after direct Fourier transformation of Eq.4 and upon separation of the polarization's vector components, we may write:

$$\tilde{P}_y=\frac{n_{0,f}e^2\lambda_0^2/(m_0^* c^2)\tilde{E}_y-\eta\tilde{k}_y\tilde{k}_z\tilde{P}_z}{\left(-\tilde{\omega}^2-i\tilde{\gamma}_f\tilde{\omega}+\frac{3}{2}\eta\tilde{k}_y^2+\frac{\eta}{2}\tilde{k}_z^2\right)}$$

$$\tilde{P}_z=\frac{n_{0,f}e^2\lambda_0^2/(m_0^* c^2)\tilde{E}_z-\eta\tilde{k}_y\tilde{k}_z\tilde{P}_y}{\left(-\tilde{\omega}^2-i\tilde{\gamma}_f\tilde{\omega}+\frac{3}{2}\eta\tilde{k}_z^2+\frac{\eta}{2}\tilde{k}_y^2\right)}\qquad \eta=\frac{3E_F}{5m_0^* c^2}\qquad \tilde{\omega}=\frac{\omega}{\omega_0}\qquad \tilde{k}_{y,z}=\lambda_0 k_{y,z}. \quad (5)$$

For typical noble metals and conductive oxides, $\eta\approx 10^{-5}$. For planar structures, each plane wave represented in Eqs.5 refracts at an angle dictated by the magnitudes of $\tilde{k}_y$ and $\tilde{k}_z$. Eqs.5 may be solved and put into the usual form: $\begin{pmatrix}\tilde{P}_y\\ \tilde{P}_z\end{pmatrix}=\begin{pmatrix}\chi_{yy} & \chi_{yz}\\ \chi_{zy} & \chi_{zz}\end{pmatrix}\begin{pmatrix}\tilde{E}_y\\ \tilde{E}_z\end{pmatrix}$. While the off-diagonal elements are generally non-zero, for uniform layers they tend to perturb the system. Therefore, for the purposes of our discussion off-diagonal elements will be neglected, in view of the relatively small magnitude of $\eta$. The form of Eqs.5 thus demonstrates that even if the medium is assumed to be isotropic via Eq.(1), nonlocal effects intervene by introducing an intrinsic anisotropy [10] that affects propagation and eventually nonlinear interactions at all angles of incidence, and as we will see below, by triggering dramatic modulation of the transverse dielectric constant. Below we examine both consequences in some detail.

Modifications of the dielectric constant due to nonlocal effects are usually understood and described almost exclusively in terms of a blueshift of the main plasmonic resonance, in this case centered near the ENZ wavelength, and by the generation of additional absorption resonances that can be correlated directly to longitudinal, resonant oscillations of the free electron gas prompted by radiation pressure [9]. In Fig.1 (a) we depict linear pump absorption spectra for 100fs, p-polarized pulses incident at a 60° angle on a 20-nm-thick ITO film suspended in vacuum, for local and nonlocal regimes. For the pump field, nonlocal effects manifest themselves primarily with the aforementioned blueshifted main peak (horizontal arrow) and additional absorption resonances, highlighted by the perpendicular arrows near 700nm and 900nm. In general, absorption cannot be calculated analytically due to the presence of dynamic pressure and viscosity terms. In Fig.1 (a) we calculate absorption as the total scattered (transmitted and reflected) pump energy subtracted from the total energy contained in the incident pulse. This approach is exact, since it is based on energy conservation.

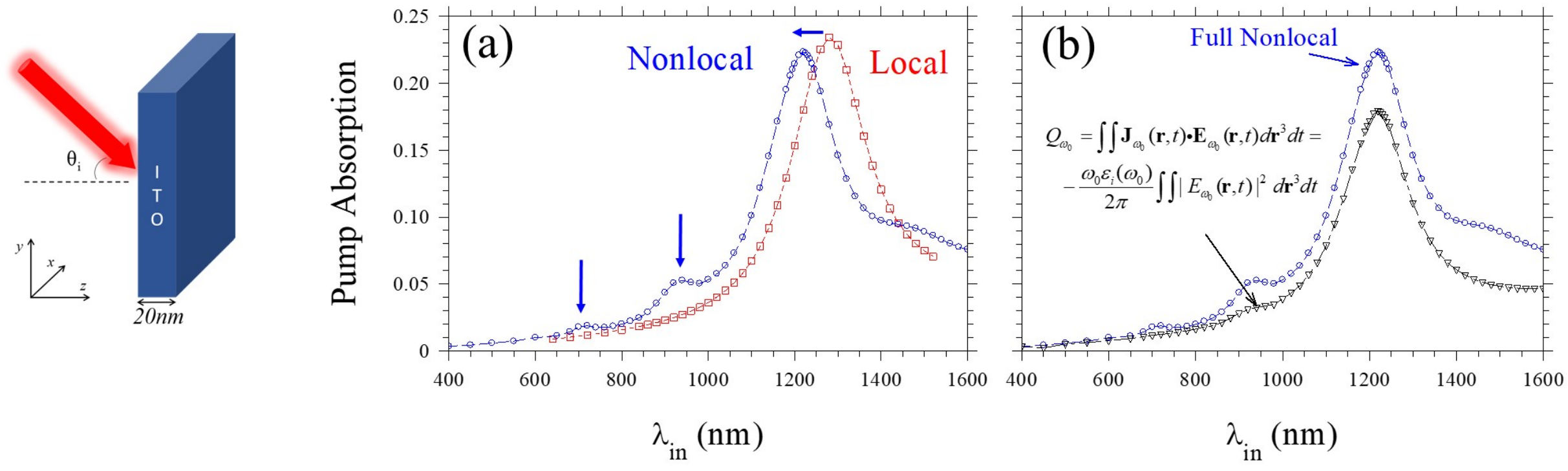

**Fig.1:** (a) Calculated local and nonlocal pump absorption spectra at 60° angle of incidence. The vertical arrows denote the locations of additional resonances that are triggered by longitudinal oscillations of the electron gas. The horizontal arrow represents a blueshift of the resonance triggered by a changing, effective dielectric constant. (b) Absorption calculated in two ways: "full nonlocal", corresponding to the curve in (a), and by using a hybrid approach consisting of using the fields calculated from the full model, and the local dielectric constant of Eq.1. This comparison illustrates that total absorption must be calculated by including the additional piece of dielectric constant triggered by nonlocal effects, a quantity that depends on the spatial derivatives of the polarization, i.e. local charge density. Left Inset: geometry of the interaction.

An aspect that is often overlooked, but is nevertheless associated with modifications of the dielectric constant, is depicted in Fig.1(b), where we compare the total, nonlocal absorption shown in Fig.1 (a) with the absorption calculated using the standard Poynting theorem, but by using the local dielectric constant. The discrepancy between the curves is obvious in both amplitude and the near-absence of additional absorption peaks, and is due to the fact that the imaginary parts of the effective susceptibilities derived from Eqs.(5) are modified in nontrivial ways. Fig.1 (b) thus strongly suggests that care should be exercised when either linear or nonlinear absorption are being considered and evaluated anytime nonlocal effects are relevant. Finally, we note that peak locations and amplitudes in Fig.1 depend on incident angle (Fig.11 below).

## 2.2 Induced Anisotropy and Reconstruction of Linear and Nonlinear Effective Dielectric Constants

We now wish to discuss a method that allows extraction of the approximate, effective linear and/or nonlinear responses of the medium under consideration, and to evaluate the intrinsic anisotropy suggested by Eqs.(5) in the general case of oblique incidence. Although the dielectric constants expressed in Eqs.(1) or (5) are never explicitly specified or introduced, they may be recovered by integrating the system comprising Eqs.(2-3) and Maxwell's equation in the time domain, and by exploiting the macroscopic constitutive relations. For instance, following the development that leads to Eqs.(5), assuming that the off-diagonal elements continue to be negligible, for a nearly monochromatic incident field we may write approximate expressions for the total polarizations: $P_y \approx \chi_{yy} E_y$, and $P_z \approx \chi_{zz} E_z$. It follows that:

$$\varepsilon_{yy} \approx 1 + 4\pi P_y / E_y \ , \qquad (6)$$

and

$$\varepsilon_{zz} \approx 1 + 4\pi P_z / E_z \, . \qquad (7)$$

It is understood that fields and polarizations are functions of position, so that both $\varepsilon_{yy}$ and $\varepsilon_{zz}$ in Eqs.(6-7) are spatially modulated by the ratio of the fields. These relations hold in both linear and nonlinear regimes, conditional on near-monochromaticity of the incident pulse. For practical purposes, a field may be said to be nearly monochromatic if its spatial extension is much larger than the structure under study, and if there are no sharp spectral features that may either span the bandwidth of the incident pulse, or that more generally may intrude in the spectral region of interest. Both conditions are satisfied for ordinary dispersive systems like a 20nm-thick ITO layer being illuminated by pulses that have a spatial extension in excess of 30 microns (~100fs in duration).

In Fig.2 we plot the complex dielectric function retrieved experimentally via spectroscopic ellipsometry (Woollam, VASE 250nm-1700nm) at multiple angles of incidence (60º-70º) for the 20nm ITO layer grown on both fused silica and silicon substrates investigated in reference [1], *purposely* fitted with the local, isotropic permittivity model of Eq.1, in order to test our propagation model. Using a Lorentz-Drude model ensures that the retrieved dielectric constant is consistent with the Kramers-Kronig relations. Also shown in Fig.2 are the effective, complex dielectric constants retrieved using Eqs.(6-7), as denoted by the labels, at low power densities (1MW/cm$^2$) and in the local approximation (no pressure and viscosity terms.) The dielectric functions are evaluated when the peak of the pulse reaches the ITO layer.

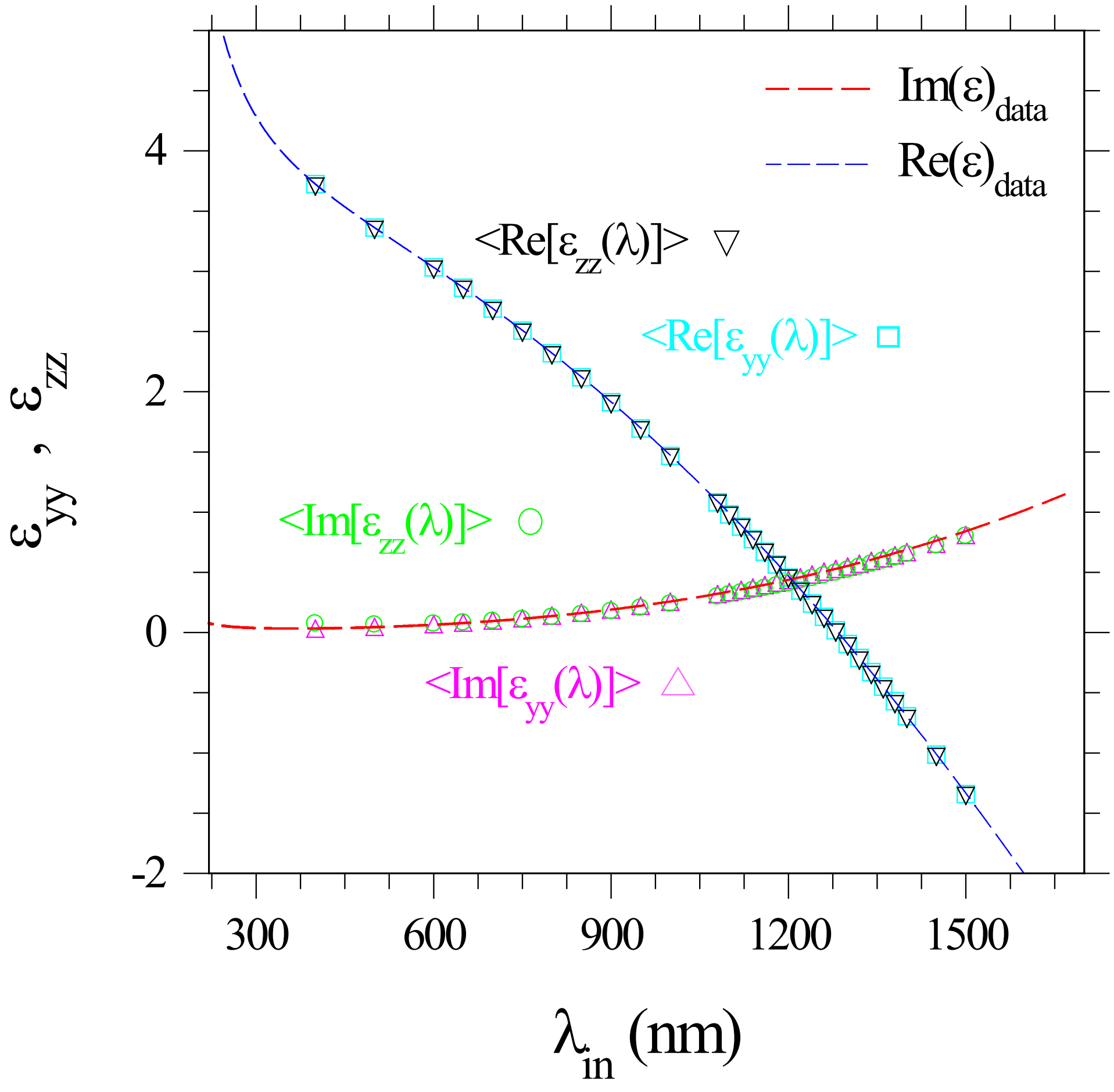

**Fig.2:** Longitudinal and transverse local dielectric constants retrieved using Eqs.(6) and (7), as indicated by the labels. The dashed curves represent our measured data purposely fitted using a local Drude-Lorentz model. The equations reproduce the data quite well over the entire range, including both Drude and Lorentz regions. In this local case the dielectric constant is isotropic.

For planar structures and arbitrary angle of incidence, the fields are uniform along the transverse coordinate, and so it suffices to perform an average along the longitudinal coordinate: $<\varepsilon_{yy,zz}(\lambda)>=\frac{1}{L}\int_0^L \varepsilon_{yy,zz}(\lambda,z)dz$, where $L$ is layer thickness. This procedure yields effective parameters and is equivalent to implementing a kind of numerical dielectric constant retrieval method on the sample. The results depicted in Fig.2 show excellent agreement between the experimentally retrieved data and our theoretical predictions, a fact that engenders confidence in our theoretical framework. Based on the results shown in Fig.2, one may also conclude that medium response is local and isotropic, i.e. $<\varepsilon_{yy}(\lambda)>=<\varepsilon_{zz}(\lambda)>$, as expected, notwithstanding the presence of the ENZ crossing point.

The introduction of nonlocal effects causes $\varepsilon_{yy}$ and $\varepsilon_{zz}$ to display unusual spatial inhomogeneities (Figs.3), while the effective dispersions $<\varepsilon_{yy}(\lambda)>$ and $<\varepsilon_{zz}(\lambda)>$ exhibit discordant ENZ crossing points (Figs.4.) Mindful of our assumptions above, in Fig.3 we plot the complex dielectric constants as functions of position via Eqs.(6-7), inside and just outside the medium, for the propagation snapshot that corresponds to the peak of the pulse reaching the ITO layer. The carrier wavelength of the pulse is ~1230nm, and the incident angle is 60°. Besides edge effects, in Fig.3 (a) $\text{Re}\left[\varepsilon_{zz}(z)\right]$ displays the expected drop to near-zero values inside the medium. Perhaps surprisingly at first, however, in Fig.3 (b) the complex $\varepsilon_{yy}(z)$ exhibits previously unreported, quite dramatic oscillatory behavior with periodicity of only a few nanometers.

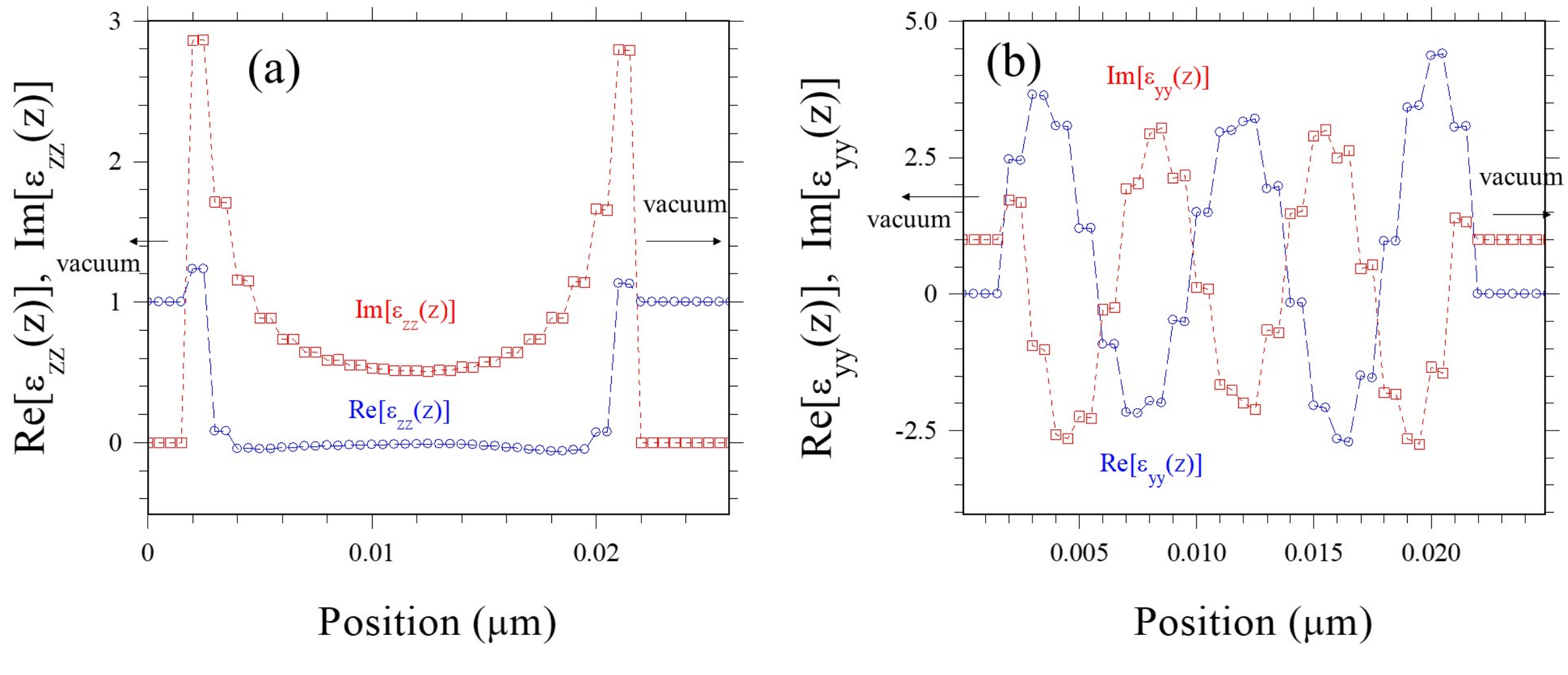

**Fig.3:** (a) Complex longitudinal and (b) transverse dielectric constants calculated using Eqs.(6) and (7). The angle of incidence is 60°. In addition to conspicuous edge effects clearly visible in (a), the imaginary component of the transverse dielectric constant takes on negative values, an indication of transient, local gain, quickly offset by local losses. The alternating sign of the effective imaginary dielectric constant is more likely an indication that local currents alternate sign within a distance of a Fermi wavelength.

Thicker layers exhibit similar oscillations only near entry and exit surfaces, accurately reflecting the fact that pressure and viscosity are felt mostly near interfaces. Even in a cursory examination of Fig.3 (b) one cannot avoid ascertaining that locally it is possible for $\text{Im}\left[\varepsilon_{yy}(z)\right]$

to be negative, which ordinarily might suggest rapid, local gain, offset by equally rapid, local loss. However, another, perhaps more physically meaningful way to view these rapid oscillations is to note that since we are dealing with mostly free electrons, nonlocal effects induce currents that alternate direction inside the layer on the scale of the Fermi wavelength, as predicted and reported for a Cadmium Oxide layer [9]. In general, the connection between conductivity and dielectric constant is easily established, and may be quantified as follows: $\sigma_{yy} = -i\omega_0 \frac{\varepsilon_{yy} - 1}{4\pi} = \frac{\omega_0}{4\pi}\left\{ \mathrm{Im}\left[\varepsilon_{yy}\right] - i\left(\mathrm{Re}\left[\varepsilon_{yy}\right] - 1\right)\right\}$, and similarly for $\sigma_{zz}$. The sign of the imaginary part thus determines the direction of local current flow. Whether or not these oscillations can ultimately be measured, possibly by probing the layer with a soft x-ray beam, is a fact presently not easily determined. However, from an effective medium standpoint, i.e. ellipsometry, their overall significance may be dismissed just as one might dismiss the significance of a phase velocity that exceeds the speed of light, an ordinary occurrence in metals. The fact is that from an effective medium standpoint, the averages $< \mathrm{Im}\left[\varepsilon_{yy,zz}(\lambda)\right] > = \frac{1}{L}\int_0^L \mathrm{Im}[\varepsilon_{yy,zz}(\lambda, z)]\,dz$ are greater than zero in all cases we have investigated.

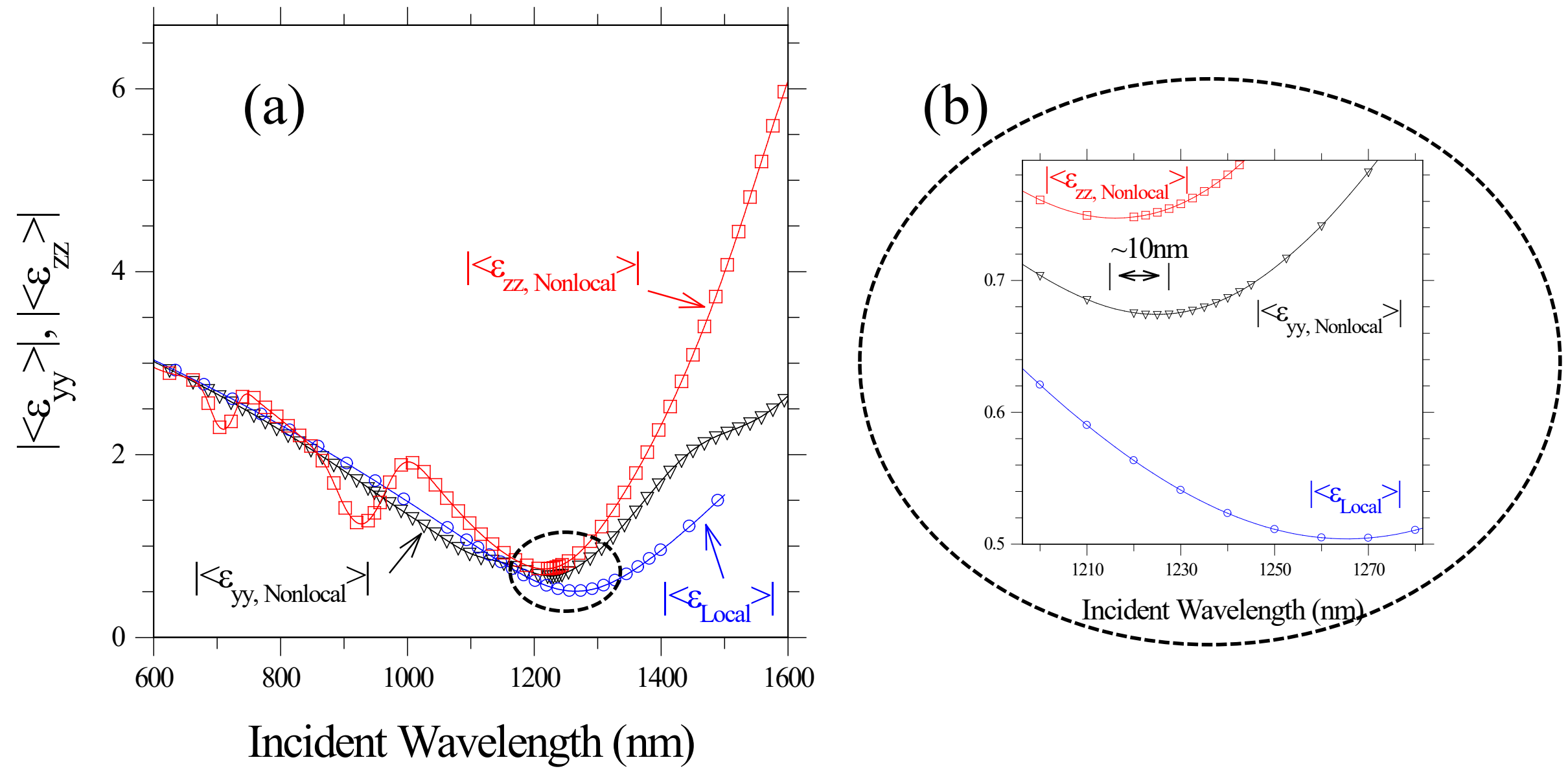

**Fig.4:** (a) Local and nonlocal longitudinal and transverse dielectric constants are averaged over the thickness of the layer. As shown in Fig.2, in the local case longitudinal and transverse dielectric constants are identical. Not so if nonlocal effects are included, which cause a shift and some degree of anisotropy, as highlighted in (b).

For illustration purposes, in Fig.4 (a) we plot only the magnitudes of the total, local and nonlocal effective dielectric constants predicted using our "numerical ellipsometry" procedure. The minimum in each curve, easily identified in Fig.4 (b), represents the ENZ crossing point.

The longitudinal dielectric constant $|<\varepsilon_{zz}(\lambda)>|$ departs most from local behavior and displays the same kind of modulation that pump absorption displays in Fig.1 (a), an indication that it drives the dynamics. In Fig.4 (b) we show a detail of the curves shown in Fig.4 (a), near the respective crossing points. In addition to an evident degree of anisotropy, the nonlocal curves are blueshifted with respect the local dielectric constant and with respect to each other, with crossing points that are mismatched by nearly 10nm. It is also important to note that at certain wavelengths the difference between $<\varepsilon_{yy}(\lambda)>$ and $<\varepsilon_{zz}(\lambda)>$ approaches zero. These points are typically referred to either as iso-index point or isotropic points, in essence wavelengths where the medium acts as if it were isotropic. As shown below, the isotropic (zero-crossing) spectral positions are intensity-dependent and undergo a spectral shift.

Before one can properly estimate how much change the dielectric constant experiences as a function of incident power density, one should first quantify how it deviates from local values when nonlocal effects are introduced. In Fig.5 we compare the magnitudes of the longitudinal effective dielectric constant $|<\varepsilon_{zz}(\lambda)>|$ in the local and nonlocal approximations, in the linear regime. The plot reveals that $|<\delta\varepsilon_{zz}>|$ can be of order unity or larger with respect to the local dielectric constant. Therefore, it is clear that assertions of demonstrations that $|<\delta\varepsilon_{zz}>|$ is of order unity necessarily require more knowledge and context than mere comparisons to the local dielectric constant.

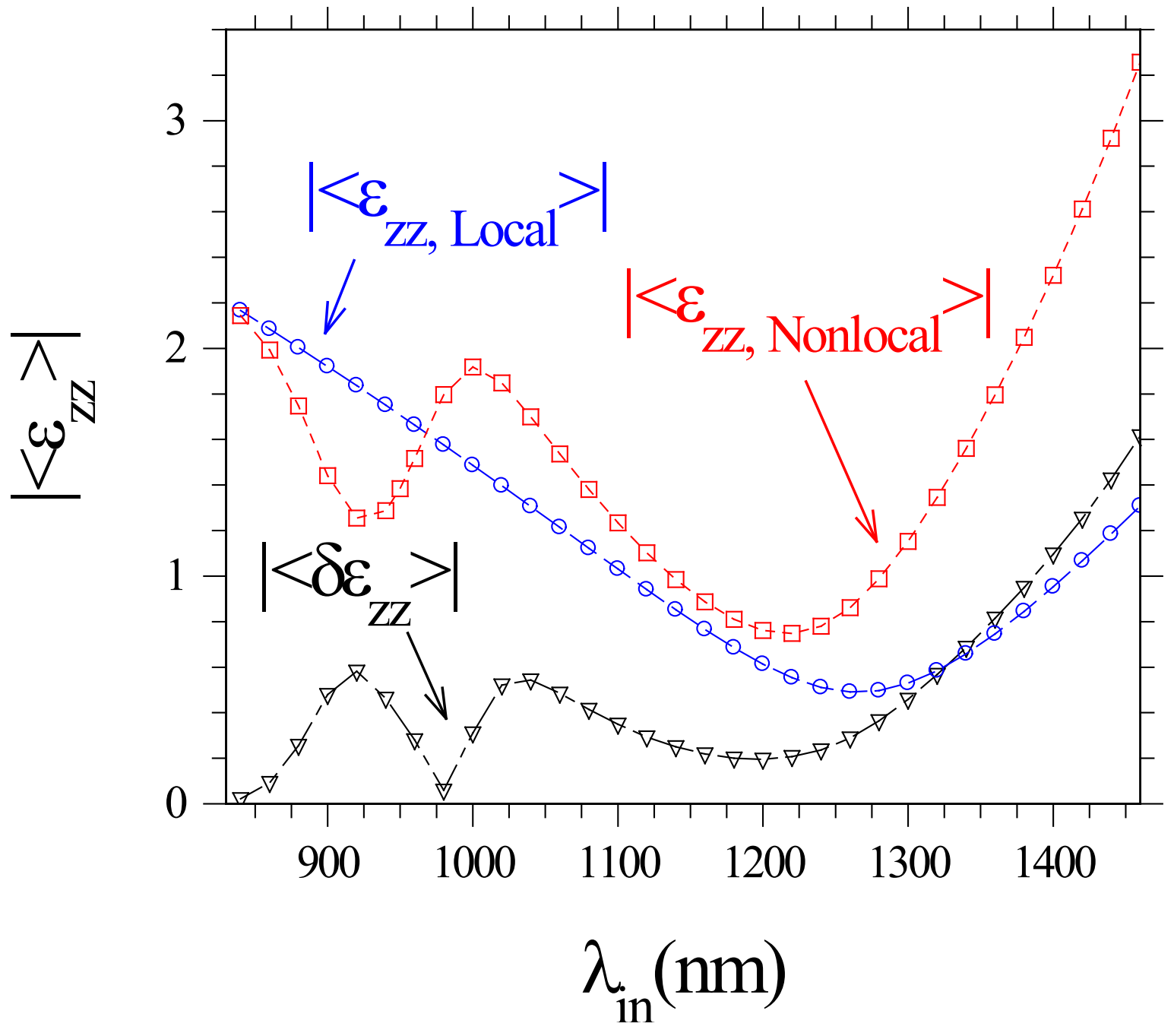

**Fig.5:** Local and nonlocal longitudinal effective dielectric constants calculated at low intensity, extracted using Eqs.(6) and (7), and the difference $|<\delta\varepsilon_{zz}>|$ between the two. The difference can be of order unity. As a result, care should be exercised when assessing the magnitude of a nonlinear index change.

Now that we have contextualized modifications of the effective dielectric constant due to nonlocal effects, we are ready to make predictions as a function of incident power density. In Fig.6 (a) we plot the amplitude of the longitudinal dielectric constant in the linear (low-intensity, 1MW/cm$^2$) and nonlinear (10GW/cm$^2$ and 20GW/cm$^2$) high intensity regimes. The curves redshift with increasing power density, following a dynamic redshift of the plasma frequency as a result of increasing effective electron mass. At these power densities, numerical stability near the ENZ conditions requires retention of hot electron nonlinearities up to 7$^{th}$ order (i.e. $\chi^{(7)}$), notwithstanding the fact that local field intensities inside the ITO layer are amplified by mere factors of two or three. Given that in the range shown the curves intersect at least in three places, the magnitude $|<\delta\varepsilon_{zz}>|$, i.e. the difference between dielectric constants in linear and nonlinear cases plotted in Fig.6 (b) approaches zero in just as many places, implying a zero index change at those locations. Therefore, it seems evident that asking oneself "How much does the index change?" may lead to ambiguous, if not misleading, answers. Instead, Figs.6 suggest more relevant questions may be asked regarding, for instance, the amount of redshift as a function of incident power density, or perhaps whether a range of constant $|<\delta\varepsilon_{zz}>|$ exists for a given incident power density. Indeed, $|<\delta\varepsilon_{zz}>|$ appears to be nearly constant in the range between 1000nm and 1200nm for both 10GW/cm$^2$ and 20GW/cm$^2$ incident power densities.

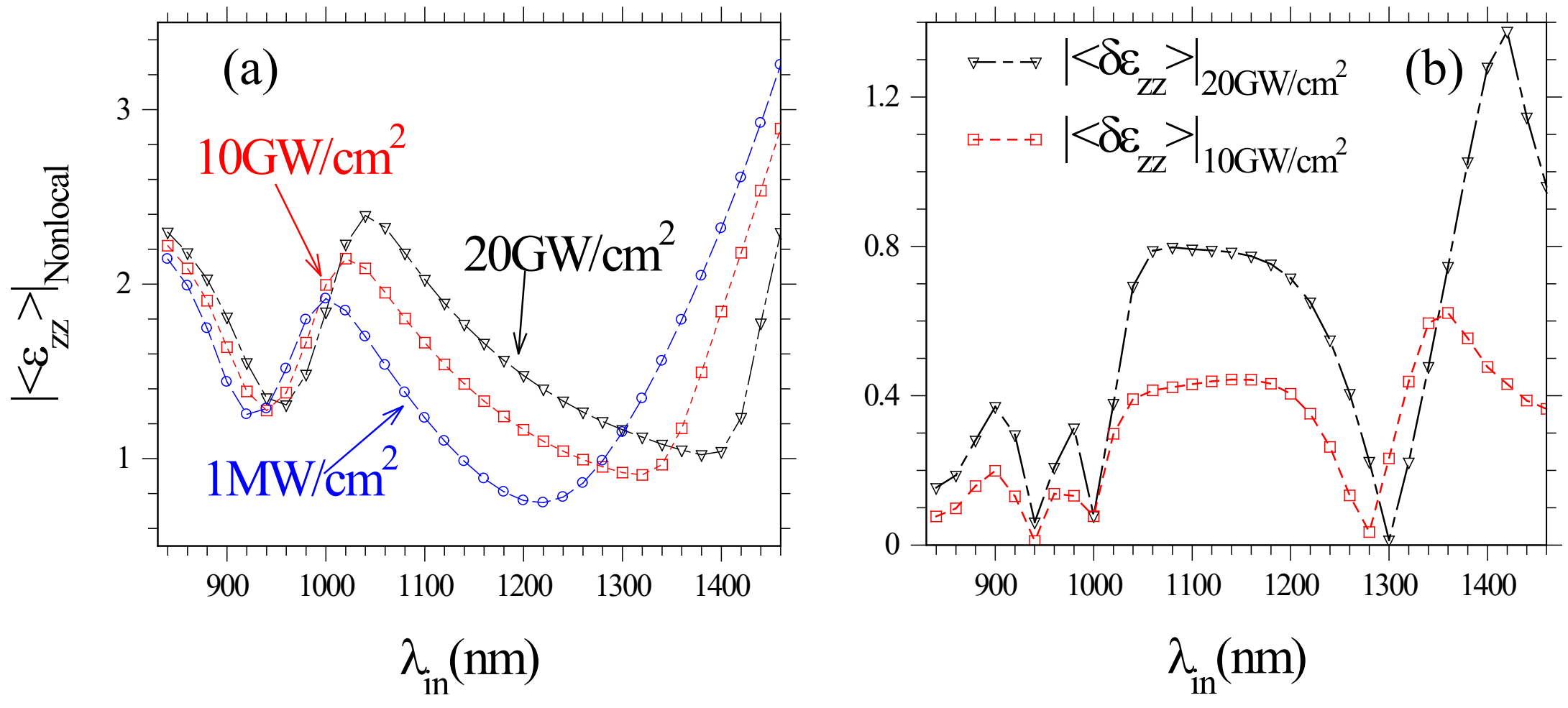

**Fig.6:** (a) Effective longitudinal dielectric constant calculated as functions of incident peak power. The redshift occurs as a result of increasing effective mass and decreasing plasma frequency. (b) Estimated magnitude of induced nonlinear change in dielectric constant $|<\delta\varepsilon_{zz}>|$ for two incident power densities with respect to the linear (or low intensity) regime. Worthy of note in (b) are that: $|<\delta\varepsilon_{zz}>|$ may be constant over a wide range of wavelengths, its magnitude can be of order unity, and for certain wavelengths $|<\delta\varepsilon_{zz}>|$ may be negligible.

Nonlinear effects may be ascertained from Fig.7, where we display several pump absorption spectra (Fig.7 (a)) as functions of incident power density. Harmonic spectra of conversion efficiencies show similar behavior. The effective, transverse and longitudinal dielectric constants are plotted in Fig.7 (b) for 20GW/cm$^2$. As alluded to above, at these power densities hot electron nonlinearities dominate the dynamics, with evident redshifts that match the redshifting dispersion curves in Fig.6 (a). In particular, the curve generated for 25GW/cm$^2$ is at the threshold for the onset of optical bistability. The discrepancies in magnitude and ENZ location between transverse and longitudinal components are evident in Fig.7 (b), and should be compared with the relatively small differences highlighted in Fig.4, in the linear regime.

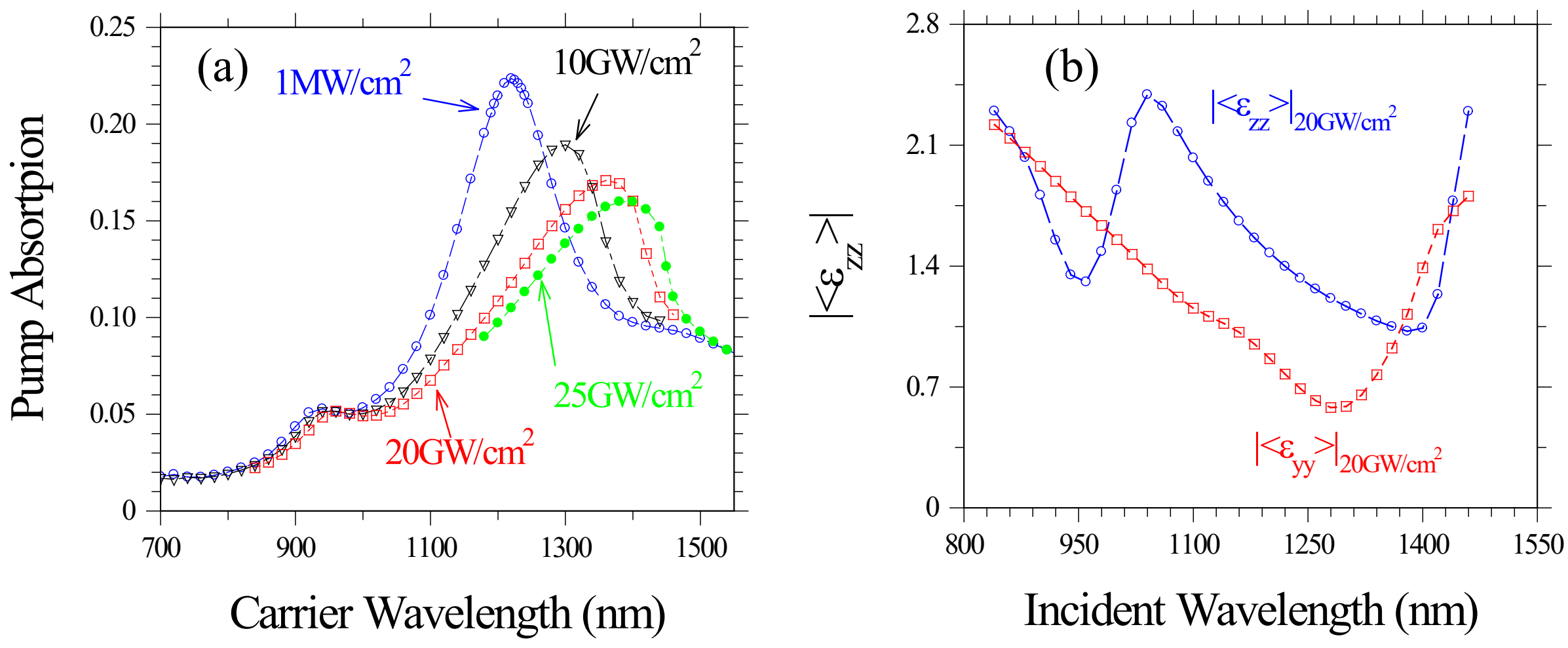

**Fig.7:** (a) Nonlinear pump absorption spectra as functions of incident peak power density. Retention of an effective seventh order nonlinearity ( $\chi^{(7)}$ ) is required to stabilize the calculation. The onset of optical bistability is imminent. (b) Effective transverse and longitudinal dielectric constants as functions of wavelength for 20GW/cm$^2$ peak power density. The absolute minima represent separate ENZ points and are ~100nm apart.

## 3. Second Harmonic Generation

Recently, experimental observations of SHG and THG were reported near 1240nm, the ENZ crossing point of a 20nm-thick ITO layer, along with predictions that employed the theoretical framework outlined above [1]. Beginning with SHG, in our present effort we expand the range of our predictions at both ends of the spectrum by extrapolating the available data, by assuming no additional factors intervene to change the dynamics, and by analyzing harmonic generation well into the ultraviolet and infrared regimes, in order to understand the interplay between free and bound electrons. Since ITO is a centrosymmetric material, second harmonic sources are found mainly in the free electron components, and consist of surface, magnetic (through the Lorentz force,) and convective terms, as outlined above. Some SHG can also come as a result of second and higher order nonlocal terms, the interaction between pump and TH photons, and are accounted in further development of Eqs.(2-3). In Fig.8 we show a comparison between reflected, local and nonlocal SHG spectra for pump wavelengths in the range 600nm-5400nm. Transmitted spectra show similar behavior. In addition to the main peak near 1240nm [1], more features are clearly predicted in both local and nonlocal responses. The maxima 1 and 2, on the blue side of the main peak, match the pump absorption resonances displayed in Fig.1 (a), triggered by longitudinal oscillations of the electron gas. All the SHG maxima above 1240nm were hitherto unknown and have different origin. The ENZ condition occurs when the pump is tuned near 1240nm, yielding a SH maximum near 620nm. A second main peak occurs for both local and nonlocal curves when the pump is tuned between 2400nm and 2500nm. Tuning the pump in that range places the SH signal near 1240nm, thus yielding a second ENZ condition, this time for the SH signal. This kind of multi-resonant enhancement has been discussed previously for a generic ENZ material [11], and for a free electron cloud that behaves as an ENZ patina that covers bulk metal layers [12, 13]. The effect can be explained along broad lines by noting that SHG efficiency is generally proportional to $1/\left(\varepsilon_\omega\sqrt{\varepsilon_{2\omega}}\right)$ [11,14]. In principle, a third peak should make an appearance when the TH signal is tuned to the ENZ conditions. However, any bulk, third order nonlinearity overwhelms that potential third resonance [12]. Finally, the last maximum located near 5000nm may be related to a maximum in the local curve that occurs near 8000nm (not shown). However, at this stage we will not pursue that wavelength range because our available data may not suffice to explain spectral features in regions that are not shown in Fig.8.

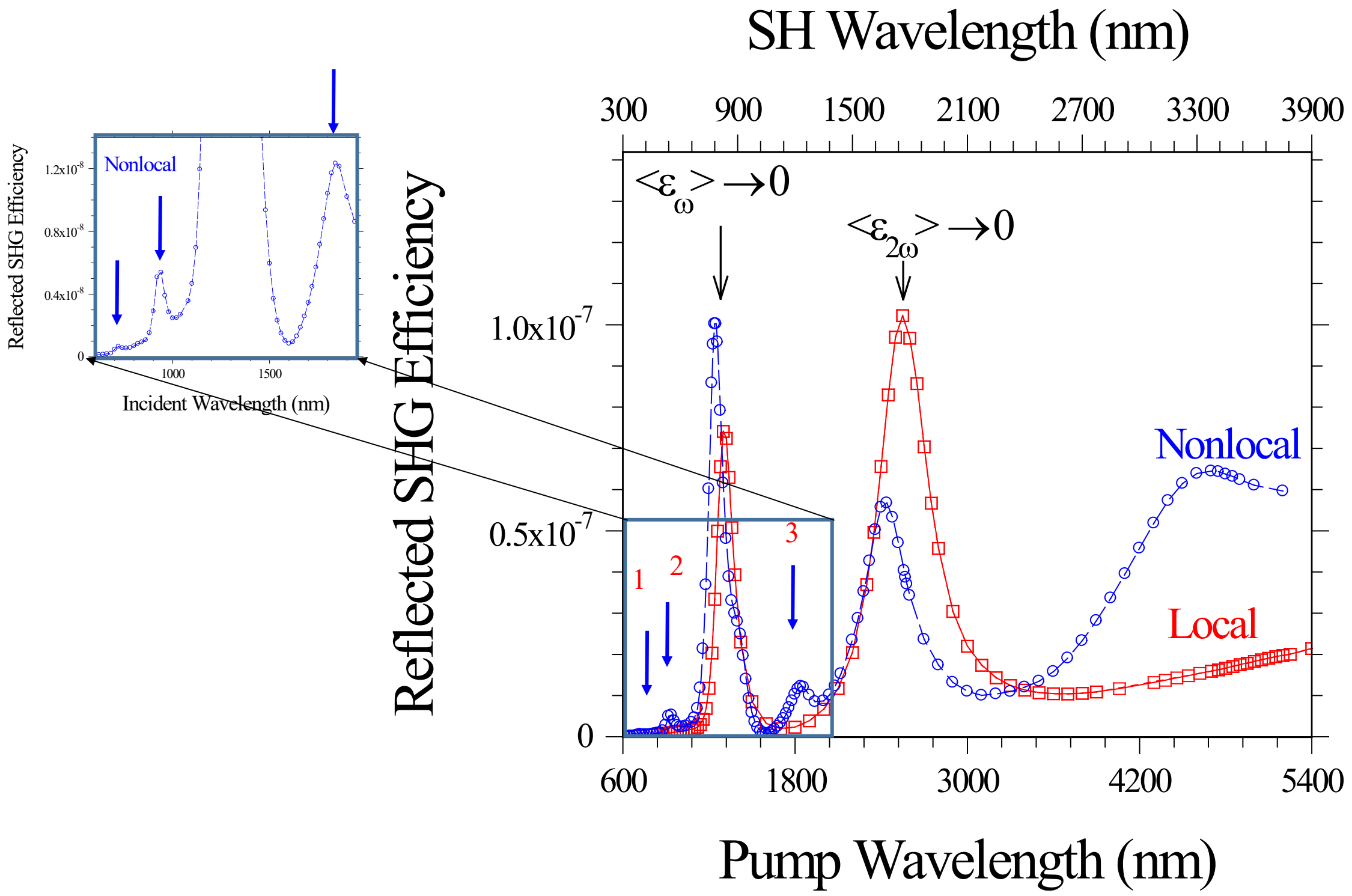

**Fig.8:** Reflected SHG conversion efficiency spectra for local and nonlocal regimes. The main peaks below 3000nm found in both local and nonlocal curves are due to respective ENZ crossing points. Additional spectral features appear in the nonlocal spectrum (inset) due to longitudinal electron gas oscillation. The local curve displays a third, broad peak near 8000nm (not shown). While nonlocal effects are clearly in play, we currently lack the necessary material data in the mid-IR range that would allow us to make definitive statements. Further studies are required to understand the nature of the nonlocal SHG peak located beyond 4000nm.

## 3. Third Harmonic Generation

Although we consider all sources of THG, including cascading from both free and bound electron, second order sources, i.e. Coulomb, Lorentz, and convective terms in both Eqs.(2) and (3), there are two main fonts of TH signal: *(i)* hot electrons, via the term $-\tilde{\Lambda}(\mathbf{E}\bullet\mathbf{E})\mathbf{E}$, and *(ii)* the bound electron nonlinear polarization component given by $\mathbf{P}_{b,NL}=\tilde{\beta}(\mathbf{P}_b\bullet\mathbf{P}_b)\mathbf{P}_b$. The temperature dependence of the free electron's effective mass may be quantified by an expression that connects linearly the effective electron mass to the electron gas temperature, i.e. $m_f^*(T_e)\approx m_e\left(0.033+aK_BT_e\right)$ [1,8], where $m_e$ is the free electron rest mass, $a$ is a constant of proportionality, $K_B$ is Boltzmann's constant, and $T_e$ is the temperature of the free electron gas that depends on absorption. Accordingly, the relative amounts of THG triggered by either nonlinear term depends on the relative amplitudes of the scaled coefficients, i.e., $\tilde{\Lambda}=aK_B\frac{n_{0,f}e^2\lambda_0^2\sigma_0\tau_0}{m_0^*c^2}$ (where $\sigma_0$ is the frequency dependent conductivity and $\tau_0$ is incident pulse duration) and $\tilde{\beta}=\frac{\omega_{0,b}^2\lambda_0^2}{L^2n_{0b}^2e^2c^2}$ (here $L$~0.3nm is the approximate lattice constant), as well as proximity to the ENZ condition and the resonant nonlinearity that is naturally elicited by the bound electrons, i.e. Lorentz oscillators.

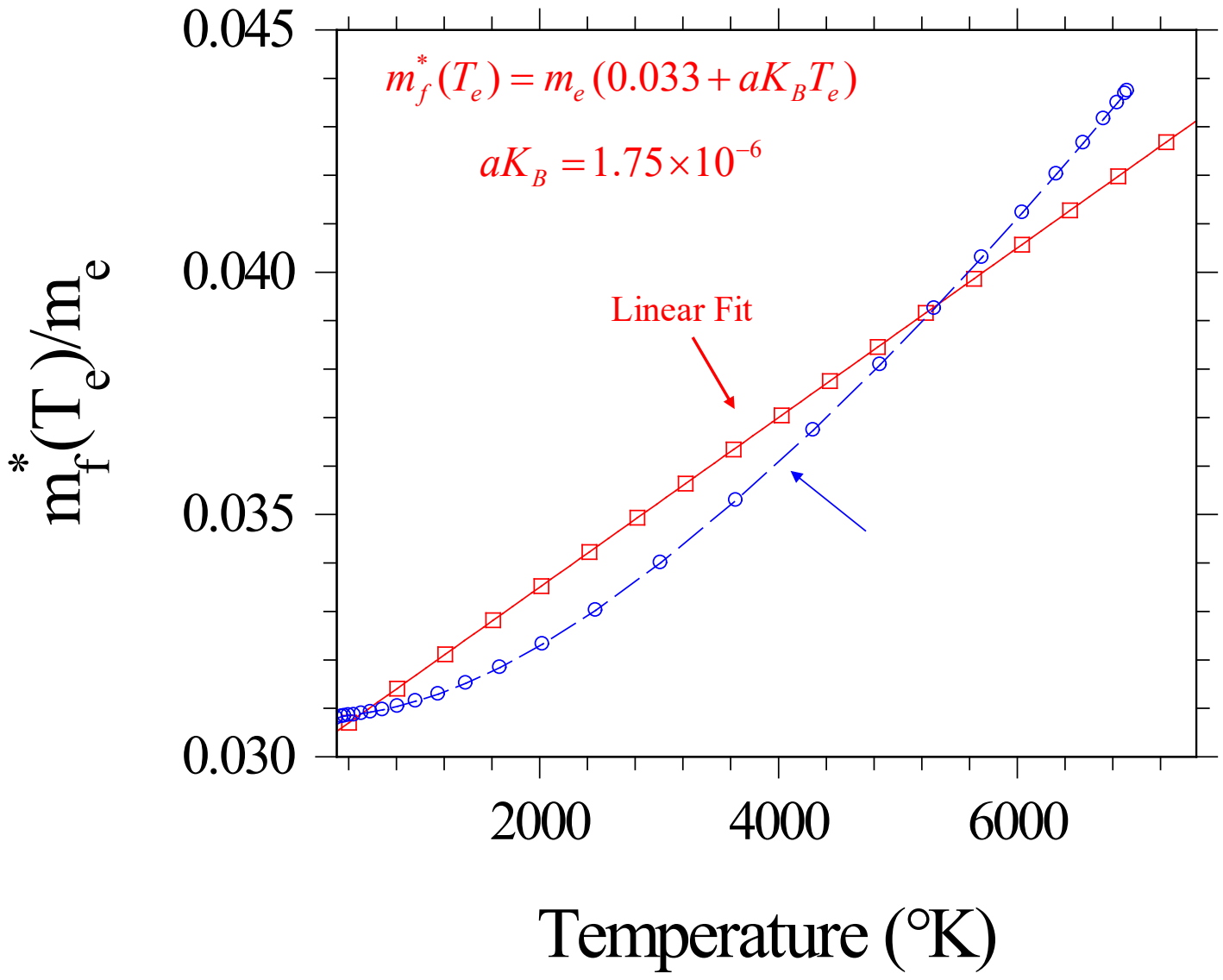

**Fig.9:** Predicted effective electron mass as a function of electron gas temperature for ITO calculated using the two-temperature model, and estimated by a linear fit, assuming $m_0=0.033m_e$.

In Fig.9 we plot the temperature dependence of the effective free electron mass extracted from the two-temperature model, following the prescription in reference [2]. The plot shows that the linear approximation we use is quite adequate, with a slope $aK_B = 1.75 \times 10^{-6}$. Based on Eq.2, the number density and the effective mass determine SH gain. Accordingly, our choices $n_0 \sim 10^{20}\,\mathrm{cm}^{-3}$ and $m_0^* \sim 0.033 m_e$ (approximate $y$-intercept) yield SHG amplitudes consistent with experimental observations for our sample [1].

Keeping in mind that parameters can be adjusted, in Fig.10 we plot our predictions for transmitted THG spectra using incident pulses approximately 100fs in duration, and peak power density of 20GW/cm$^2$. Reflection curves display similar behavior. Curve (a) corresponds to the pump absorption curve in Fig.7 having the same incident peak power density, and reflects approximate nominal values associated with our linear fit of the temperature dependence of the effective mass and lattice constant, as outlined above. Although not shown, the dielectric constant in Fig.2 displays an absorption resonance near 150 nm, with a corresponding resonant nonlinearity capable of significantly enhancing THG even under condition of high, nominal absorption [15]. In this case, the hot electron nonlinearity clearly dominates. The TH peak that arises from the Lorentzian portion of the dielectric response resonance is identified by the blue arrow, and is one order of magnitude smaller compared to the TH originating at the ENZ resonance. The spectrum exemplified by curve (b) is obtained by artificially increasing $\tilde{\beta}$ by one order of magnitude compared to its value in curve (a), so that we may ascertain the relative impact of the bound electron resonance with respect to the hot electron contribution. This increase translates to a two-order of magnitude increase in conversion efficiency at the bound electron resonance, which now dominates over the TH signal originating at the ENZ peak. This peak displays a much more modest increase of conversion efficiency compared to curve (a) because it is located in the evanescent region of the Lorentz resonance. Finally, in curve (c) we use the parameters of curve (b) and set $\tilde{\Lambda} = 0$, equivalent to turning off the hot electron contribution. These results suggest that in the high intensity regime the two types of nonlinearities may be identified, and that hot electrons govern THG by shifting and distorting the ENZ resonance, and by contributing little near the Lorentz resonance.

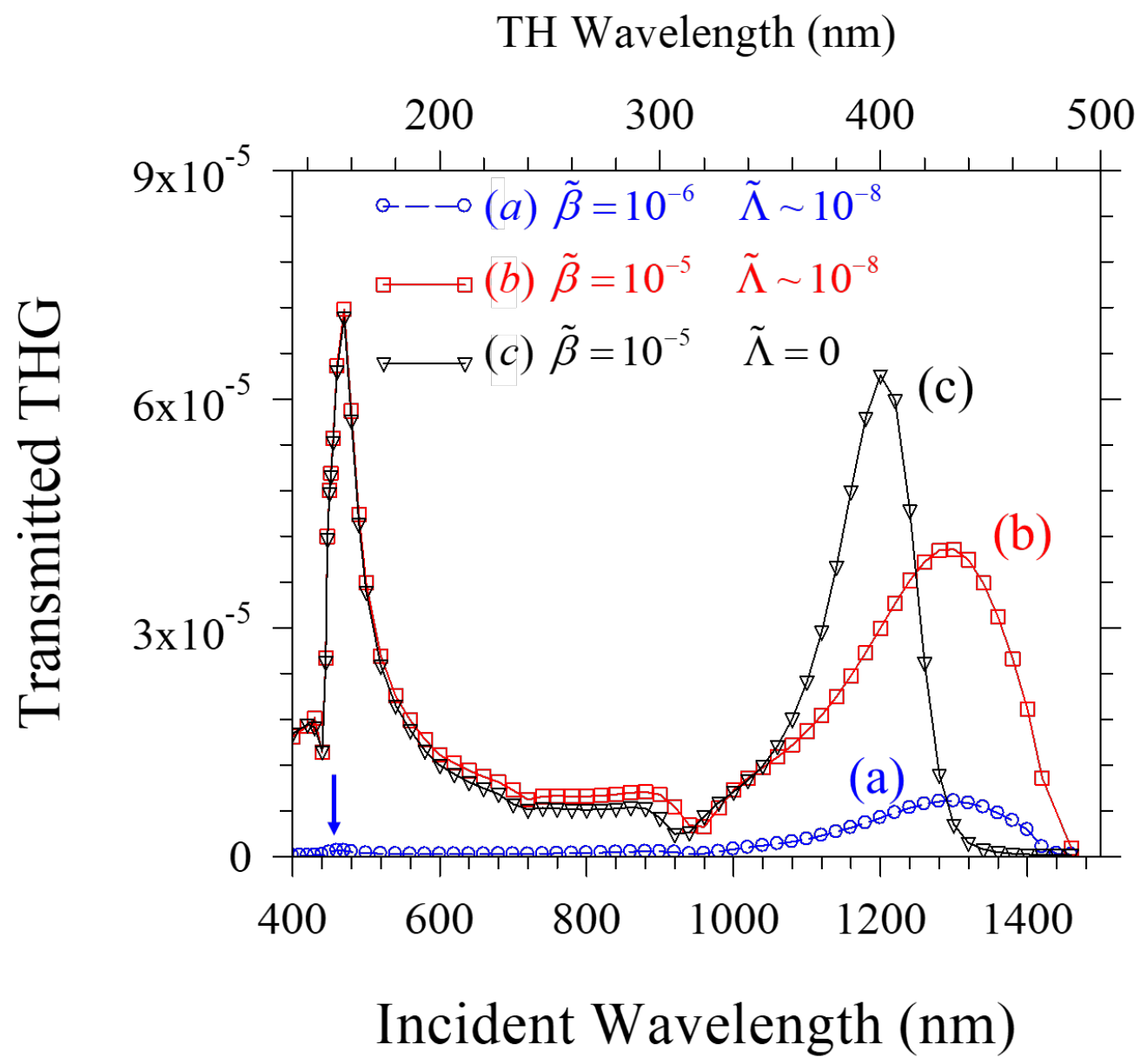

**Fig.10:** Predicted transmitted THG conversion efficiency spectra as functions of scaled third order coefficients. The magnitudes and relative amplitudes of $\tilde{\beta}$ and $\tilde{\Lambda}$ determine the range of influence of each of these components.

## 4. Additional Considerations and General Aspects of the Model beyond ITO

The sensitivity of the dynamics to incident angle may be ascertained from Fig.11, where we display both pump absorption and transmitted THG spectra for three incident angles in the full, nonlocal regime. The figures clearly suggest that retrieving the effective dielectric constant should be done carefully, with the stipulation that experimental characterization of each sample should be coupled with in-depth numerical analysis of the retrieved functions.

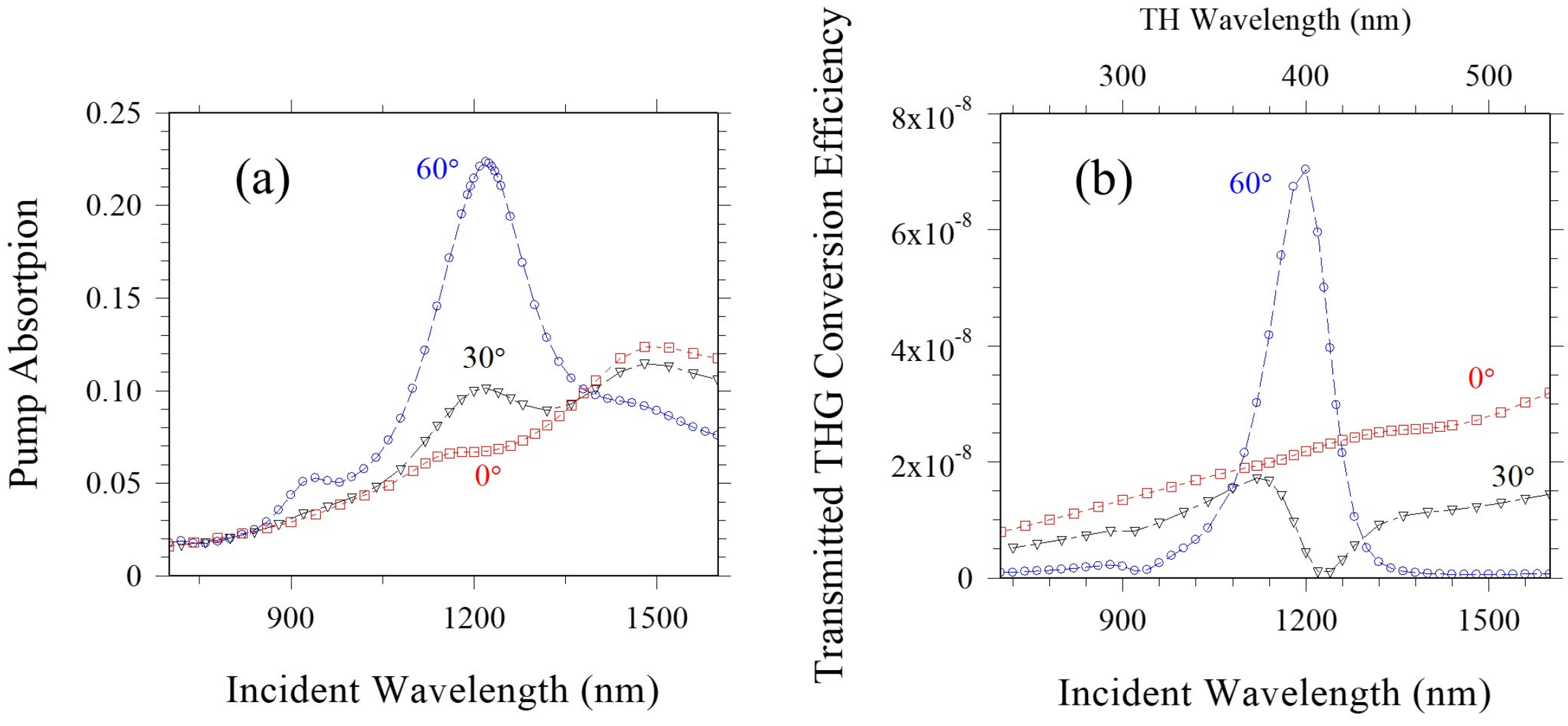

**Fig.11:** (a) Pump absorption and (b) transmitted THG spectra for three different incident angles, as indicated by the labels. Reflected THG displays similar behavior. The sensitivity to incident angle is particularly obvious in the THG spectra, where both resonant (60°) and anti-resonant (30°) behavior can be excited. The minimum predicted at 30° has been experimentally observed and reported in Reference [1].

An issue that will be addressed in further developments of the model is the apparently instantaneous nature of the hot electron dynamics in Eq.2. The two-temperature model (TTM) imparts a temporal delay that may to some extent affect the dynamics. A possible, simple cure for this shortcoming is to perform the temporal integral of absorption when estimating the effective mass. While an examination of Fig.9 reveals that the mass may be assumed to vary approximately linearly with temperature, the curve obtained by applying the TTM may be faithfully reproduced by a polynomial function of temperature. In our case an expansion to second order suffices: $m_f^*(T_e) \approx m_0^*\left(1 + aT_e + bT_e^2\right)$, where, as noted in [1], temperature is proportional to fluence: $T_e(t) = \Lambda \int_{-\infty}^{t} \dot{\mathbf{P}}_f \bullet \mathbf{E} dt'$. This formulation implies that the TTM may need to be used only once at the start in order to estimate the mass dependence with temperature for a given geometry, followed by fitting the mass with a simple polynomial in powers of the temperature, and finally by using the free electron current density, $\dot{\mathbf{P}}_f$, calculated from Eq.2 in order to automatically include nonlocal effects.

The validity of Eq.2 above is predicated on the fact that the total number of free charges within a specified volume remains constant, subject to the continuity equation: $\dot{n}_f(\mathbf{r},t) = -\frac{1}{e}\nabla \bullet \dot{\mathbf{P}}_f(\mathbf{r},t)$. However, in both metals and semiconductors the free charge density can change as a result of exciting valence electrons into the conduction band. Under these circumstances, the continuity equation may be modified to include a source term as follows:

$$\dot{n} = -\frac{1}{e}\nabla \bullet \dot{\mathbf{P}}_f + \Omega\frac{\partial}{\partial t}\mathbf{E} \bullet \mathbf{E}. \qquad (8)$$

$\Omega$ is a proportionality constant. Eq.(8) may be integrated directly to yield:

$$n = n_0 - \frac{1}{e}\nabla \bullet \mathbf{P}_f + \Omega \mathbf{E} \bullet \mathbf{E}. \qquad (9)$$

Spatial and temporal dependence are implied but left out for simplicity. These modifications thus allow the free electron number density to increase with increasing peak power density, and must be applied to the following, unscaled free-electron equation of motion [16]:

$$\ddot{\mathbf{P}}_f - \frac{\dot{n}}{n}\dot{\mathbf{P}}_f + \left(\dot{\mathbf{P}}_f \bullet \nabla\right)\left(\frac{\dot{\mathbf{P}}_f}{ne}\right) + \gamma\dot{\mathbf{P}}_f = \frac{ne^2}{m}\mathbf{E} + \frac{e}{mc}\dot{\mathbf{P}}_f \times \mathbf{B} - \frac{e\nabla p}{m}, \qquad (10)$$

where we have retained only the nonlocal pressure term to illustrate modifications that Eqs.(8) and (9) induce on nonlocal effects. Eq.2 may be derived directly from Eq.(10) when $\Omega = 0$. In the general case, the number density and the effective mass can undergo changes, especially at high intensities. However, for simplicity we will neglect effective mass changes, which are represented in Eq.2. Using Eqs.(8-9) and the expression that describes quantum pressure, $p = p_0\left(\frac{n}{n_0}\right)^{5/3}$, with $p_0 = n_0 E_F$ the new, scaled Eq.(2) takes the following form:

$$\ddot{\mathbf{P}}_f + \tilde{\gamma}_f \dot{\mathbf{P}}_f = \frac{n_{0,f} e^2 \lambda_0^2}{m_0^* c^2}\mathbf{E} - \frac{e\,\lambda_0}{m_0^* c^2}\mathbf{E}\left(\nabla \bullet \mathbf{P}_f\right) + \frac{\Omega e^2 \lambda_0^2}{m_0^* c^2}\left(\mathbf{E}\bullet\mathbf{E}\right)\mathbf{E} + \frac{e\,\lambda_0}{m_0^* c^2}\dot{\mathbf{P}}_f \times \mathbf{H}$$
$$-\frac{1}{n_{0,f} e \lambda_0}\left[\left(\nabla \bullet \dot{\mathbf{P}}_f\right)\dot{\mathbf{P}}_f + \left(\dot{\mathbf{P}}_f \bullet \nabla\right)\dot{\mathbf{P}}_f\right] \qquad . \qquad (11)$$
$$+\frac{5}{3}\frac{E_F}{m_0^* c^2}\left(\nabla\left(\nabla \bullet \mathbf{P}_f\right) - \frac{2}{3 e n_0 \lambda_0}\left(\nabla \bullet \mathbf{P}_f\right)\nabla\left(\nabla \bullet \mathbf{P}_f\right) - 2\Omega e \lambda_0 \left(\mathbf{E}\bullet\nabla\right)\mathbf{E}\right) + \frac{\Omega}{n_0}\dot{\mathbf{P}}_f \frac{\partial}{\partial \tau}\left(\mathbf{E}\bullet\mathbf{E}\right)$$

The clear outcome of allowing interband transitions is a blueshifted, intensity-dependent plasma frequency, while convection and nonlocality produce supplementary nonlinear terms that represent new second order surface $\left(-\frac{10}{9}\frac{E_F}{m_0^* c^2}\frac{1}{3 e n_0 \lambda_0}\left(\nabla \bullet \mathbf{P}\right)\nabla\left(\nabla \bullet \mathbf{P}\right) - \frac{10}{3}\frac{e\Omega\lambda_0 E_F}{m_0^* c^2}\left(\mathbf{E}\bullet\nabla\right)\mathbf{E}\right)$ and third order volume $\left(\frac{\Omega}{n_0}\dot{\mathbf{P}}_f \frac{\partial}{\partial \tau}\left(\mathbf{E}\bullet\mathbf{E}\right)\right)$ sources, with consequences for SHG and THG. Future work will include an assessment of the new material equation of motion (11) as it may apply to metals and semiconductors.

## 5. Conclusions and Summary

We have discussed a pulse propagation model that accounts for hot carriers, pump depletion, surface and volume nonlinear sources, as well as free and bound electron contributions in the context of a 20nm-thick ITO layer. We have predicted that nonlocal effects induce anisotropic medium response, and that at high enough intensities optical bistability is triggered. The method can be used to retrieve effective dielectric response in both linear and nonlinear regimes, making it possible to predict the amount of nonlinear index or dielectric change as functions of incident power density. At sufficiently large intensities, we are able to discriminate between third order free and bound electron contributions to THG. We also predict previously unknown spectral features of SHG, partly due to nonlocal effects, and in part arising from a SH signal tuned to the ENZ condition. The implication of the induced anisotropy by nonlocal effects is twofold: on one hand it modifies the linear response and the propagation inside the medium as shown above, on the other, it may provide additional tools to tune and enhance nonlinear phenomena like harmonic generation [17]. Finally, the model can also account for interband transitions that drive valence electrons into the conduction band, leading to increased free change density and additional second and third order nonlinear sources that may help to shed light on a host of nanoscale phenomena in both metals and semiconductors.

**Acknowledgments**

MS acknowledges discussions Antonino Cala' Lesina. MAV acknowledges financial support from the Rita Levi-Montalcini Italian Ministry of Education and Research. LRS, JT and CC acknowledge financial support from RDECOM Grant W911NF-18-1-0126 from the International Technology Center-Atlantic. Research of D. d. C. was performed within Project "Internet of Things: Sviluppi Metodologici, Tecnologici E Applicativi", co-funded (2018-2022) by the Italian Ministry of Education, Universities and Research (MIUR) under the aegis of the "Fondo per il finanziamento dei dipartimenti universitari di eccellenza" initiative (Law 232/2016).